\documentclass[]{spie}  

\usepackage{amsmath,amsfonts,amssymb}
\usepackage{graphicx}
\usepackage[colorlinks=true, allcolors=blue]{hyperref}

\title{Development status of the SOXS spectrograph for the ESO-NTT telescope}

\author[a]{P.~Schipani}
\author[b]{S.~Campana}
\author[c]{R.~Claudi}
\author[b]{M.~Aliverti}
\author[a]{A.~Baruffolo}
\author[d]{S.~Ben-Ami}
\author[c,v]{F.~Biondi}
\author[a]{G.~Capasso}
\author[e]{R.~Cosentino}
\author[f]{F.~D'Alessio}
\author[b]{P.~D'Avanzo}
\author[d]{O.~Hershko}
\author[g,h]{H.~Kuncarayakti}
\author[b]{M.~Landoni}
\author[k]{M.~Munari}
\author[m,n]{G.~Pignata}
\author[o,d]{A.~Rubin}
\author[k]{S.~Scuderi}
\author[f]{F.~Vitali}
\author[p]{D.~Young}
\author[l]{J.~Achrén}
\author[r,n]{J.~A.~Araiza-Duran}
\author[s]{I.~Arcavi}
\author[m,z]{A.~Brucalassi}
\author[d]{R.~Bruch}
\author[c]{E.~Cappellaro}
\author[a]{M.~Colapietro}
\author[a]{M.~Della~Valle}
\author[c]{M.~De~Pascale}
\author[k]{R.~Di~Benedetto}
\author[a]{S.~D'Orsi}
\author[d]{A.~Gal-Yam}
\author[b]{M.~Genoni}
\author[e]{M.~Hernandez}
\author[h,g]{J.~Kotilainen}
\author[t]{G.~Li~Causi}
\author[a]{L.~Marty}
\author[g]{S.~Mattila}
\author[c]{K.~Radhakrishnan}
\author[d]{M.~Rappaport}
\author[a]{D.~Ricci}
\author[b]{M.~Riva}
\author[c]{B.~Salasnich}
\author[a]{S.~Savarese}
\author[p]{S.~Smartt}
\author[k]{R.~Zanmar~Sanchez}
\author[u]{M.~Stritzinger}
\author[e]{H.~Ventura}
\author[o]{L.~Pasquini}
\author[o]{M.~Sch\"oller}
\author[o]{H.-U.~K\"aufl}
\author[o]{M.~Accardo}
\author[o]{L.~Mehrgan}
\author[o]{E.~Pompei}

\affil[a]{INAF -- Osservatorio Astronomico di Capodimonte, Sal. Moiariello 16, I-80131, Naples, Italy }
\affil[b]{INAF -- Osservatorio Astronomico di Brera, Via Bianchi 46, I-23807, Merate, Italy }
\affil[c]{INAF -- Osservatorio Astronomico di Padova, Vicolo dell’Osservatorio 5, I-35122, Padua, Italy }
\affil[d]{Weizmann Institute of Science, Herzl St 234, Rehovot, 7610001, Israel }
\affil[e]{FGG-INAF, TNG, Rambla J.A. Fernández Pérez 7, E-38712 Breña Baja (TF), Spain }
\affil[f]{INAF -- Osservatorio Astronomico di Roma, Via Frascati 33, I-00078 M. Porzio Catone, Italy }
\affil[g]{Tuorla Observatory, Dept. of Physics and Astronomy, University of Turku,  FI-20014, Finland }\affil[h]{Finnish Centre for Astronomy with ESO (FINCA), FI-20014 University of Turku, Finland }
\affil[k]{INAF -- Osservatorio Astrofisico di Catania, Via S. Sofia 78 30, I-95123 Catania, Italy }
\affil[m]{Universidad Andres Bello, Avda. Republica 252, Santiago, Chile }
\affil[n]{MAS, Nuncio Monseñor Sotero Sanz 100, Providencia, Santiago, Chile }
\affil[o]{ESO, Karl Schwarzschild Strasse 2, D-85748, Garching bei München, Germany }
\affil[p]{Astrophysics Research Centre, Queen's University Belfast, Belfast, BT7 1NN, UK }
\affil[q]{Incident Angle Oy, Capsiankatu 4 A 29, FI-20320 Turku, Finland }
\affil[r]{Centro de Investigaciones en Optica A. C., 37150 León, Mexico }
\affil[s]{Tel Aviv University, Department of Astrophysics, 69978 Tel Aviv, Israel }
\affil[t]{INAF - Istituto di Astrofisica e Planetologia Spaziali, Rome, Italy}
\affil[u]{Aarhus University, Ny Munkegade 120, D-8000 Aarhus, Denmark }
\affil[v]{MPE, Giessenbachstr. 1, D-85748 Garching, Germany}
\affil[z]{INAF - Osservatorio Astrofisico di Arcetri, Largo E. Fermi 5, 50125, Firenze, Italy}

\authorinfo{Send correspondence to: pietro.schipani@inaf.it}

\begin{document} 
\maketitle

\begin{abstract}
SOXS (Son Of X-Shooter) is a single object spectrograph, characterized by offering a wide simultaneous spectral coverage from U- to H-band, built by an international consortium for the 3.6-m ESO New Technology Telescope at the La Silla Observatory, in the Southern part of the Chilean Atacama Desert. The consortium is focussed on a clear scientific goal: the spectrograph will observe all kind of transient and variable sources discovered by different surveys with a highly flexible schedule, updated daily, based on the Target of Opportunity concept. It will provide a key spectroscopic partner to any kind of imaging survey, becoming one of the premier transient follow-up instruments in the Southern hemisphere.
SOXS will study a mixture of transients encompassing all distance scales and branches of astronomy, including fast alerts (such as gamma-ray bursts and gravitational waves), mid-term alerts (such as supernovae and X-ray transients), and fixed-time events (such as the close-by passage of a minor planet or exoplanets). It will also have the scope to observe active galactic nuclei and blazars, tidal disruption events, fast radio bursts, and more. Besides of the consortium programs on guaranteed time, the instrument is offered to the ESO community for any kind of astrophysical target. The project has passed the Final Design Review and is currently in manufacturing and integration phase. This paper describes the development status of the project.
\end{abstract}


\begin{figure} [ht]
\begin{center}
\begin{tabular}{c} 
\includegraphics[height=11.9cm]{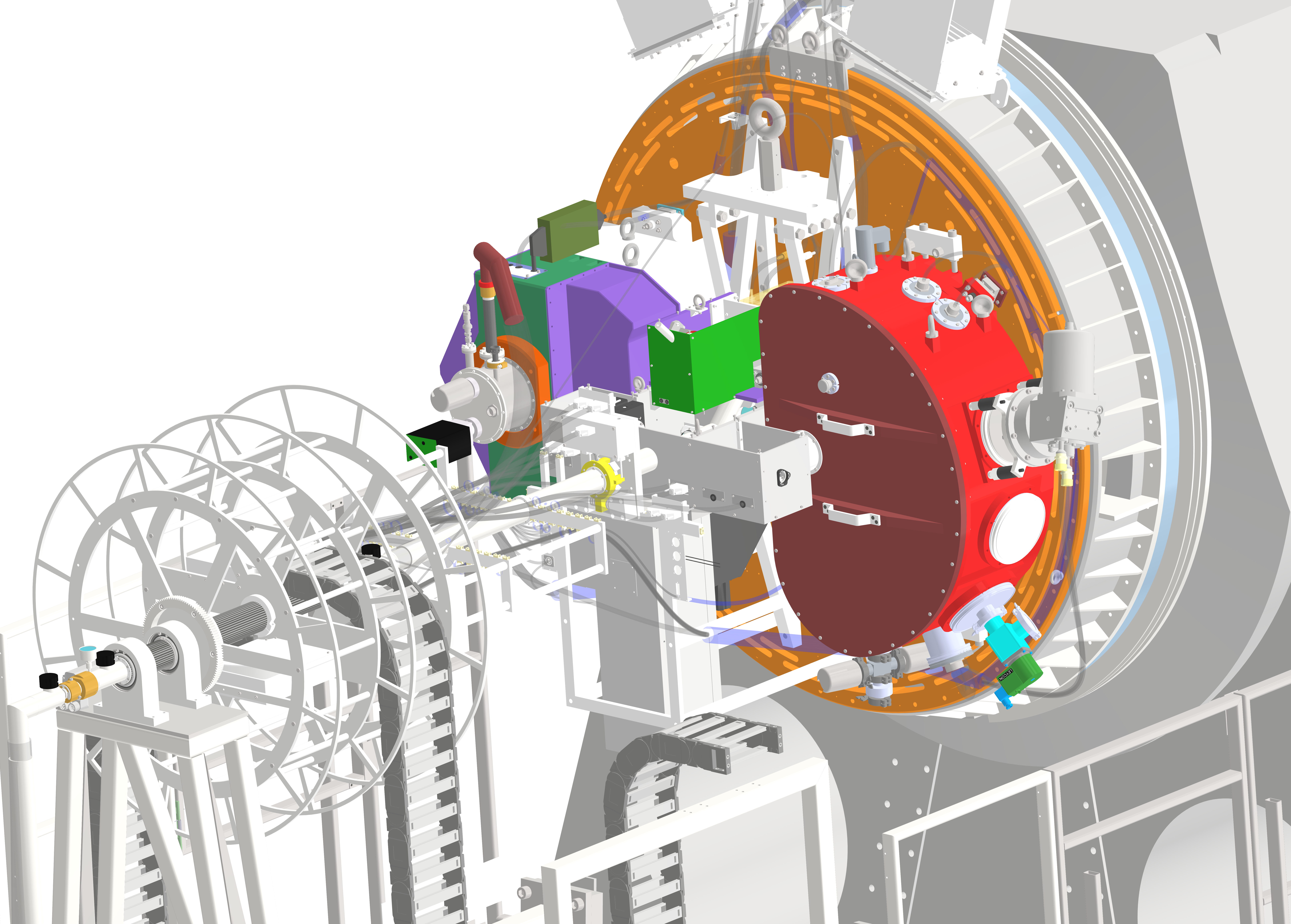}
\end{tabular}
\end{center}
\caption[example1] 
{ \label{fig:SOXS} 
SOXS at the Nasmyth focus of the NTT.}
\end{figure}

\begin{figure} [ht]
\begin{center}
\begin{tabular}{c} 
\includegraphics[height=12cm]{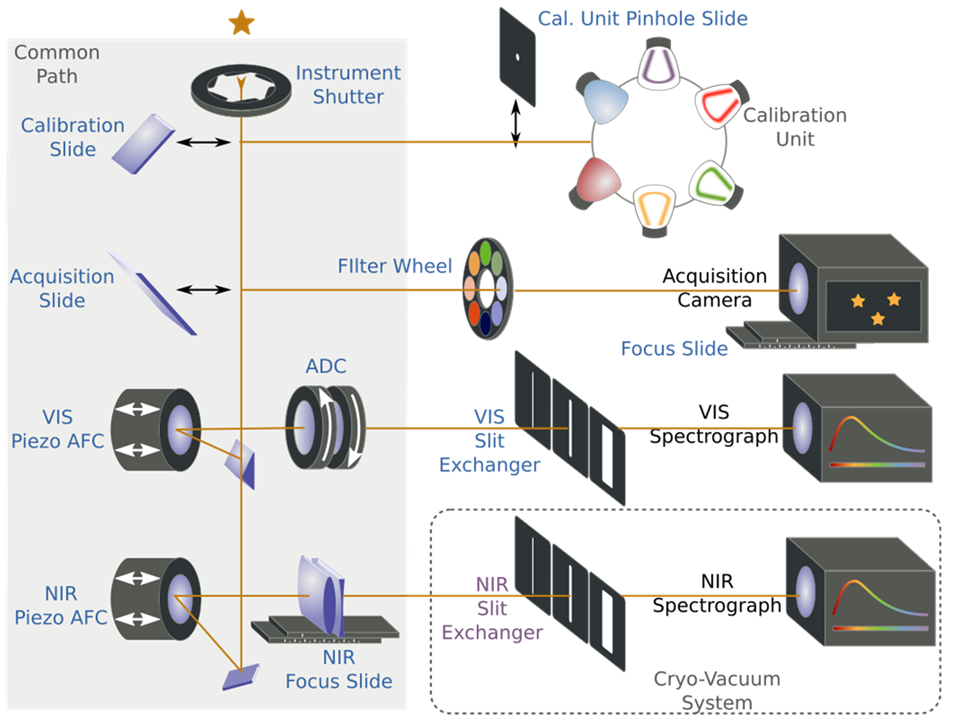}
\end{tabular}
\end{center}
\caption[example1] 
{ \label{fig:control} 
The SOXS functional diagram.}
\end{figure}

\begin{figure}
\begin{center}
  \includegraphics[height=9.9cm]{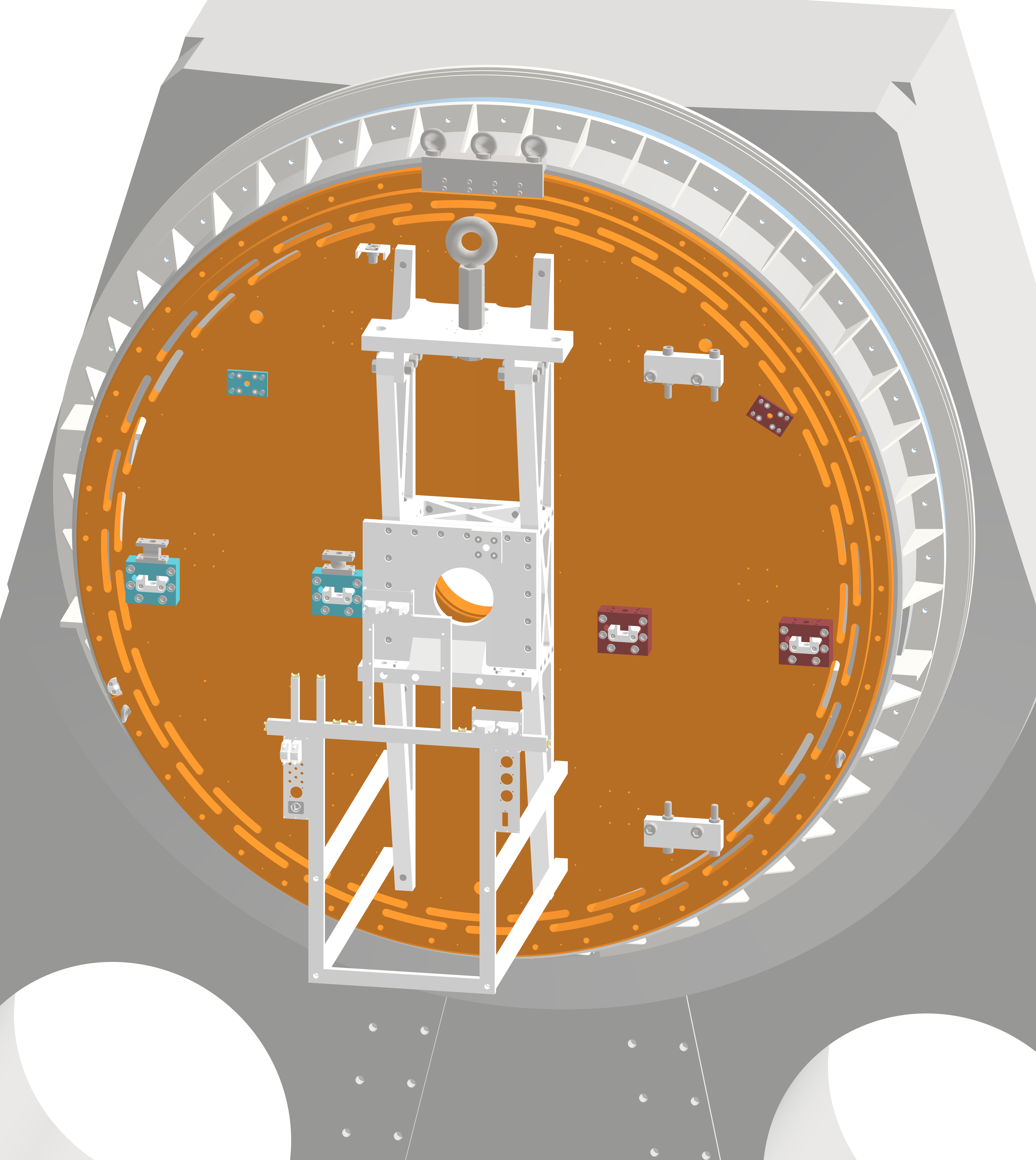}
  \includegraphics[height=9.9cm]{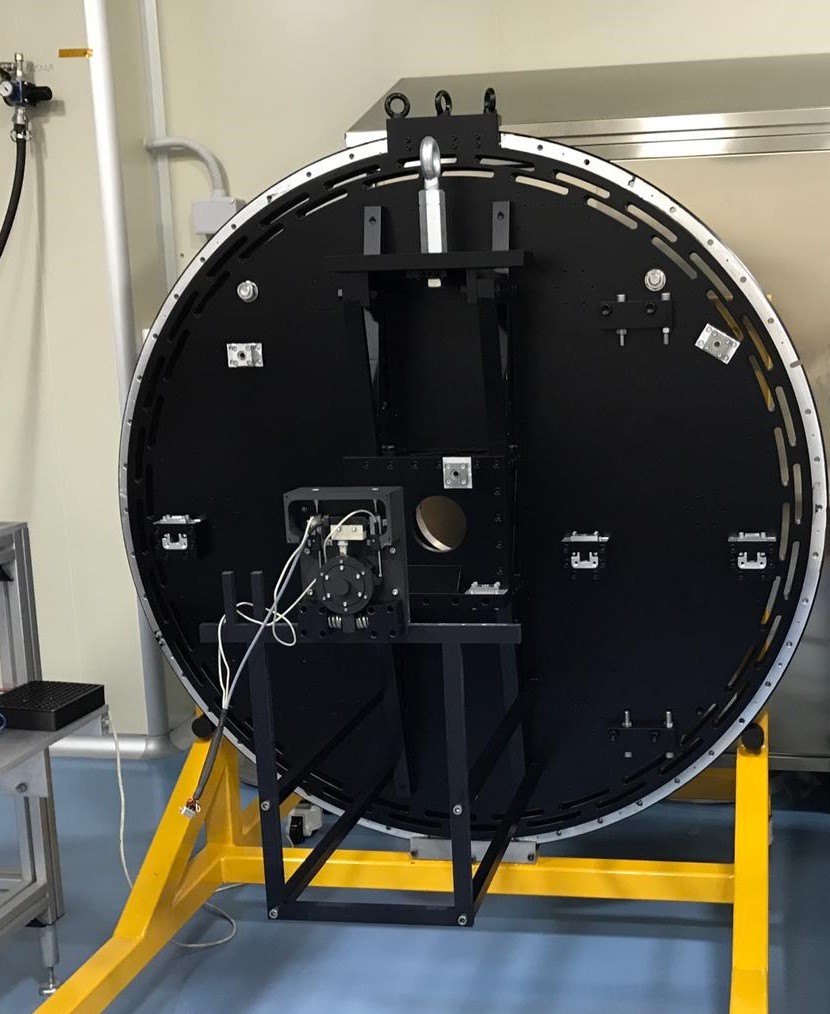}
\end{center}
\caption[example1] 
{ \label{fig:flange} 
The interface flange with the Nasmyth rotator equipped with the co-rotator feedback system. Left: design; Right: as built.}
\end{figure}

\begin{figure} [ht]
\begin{center}
\begin{tabular}{c} 
\includegraphics[height=7.5cm]{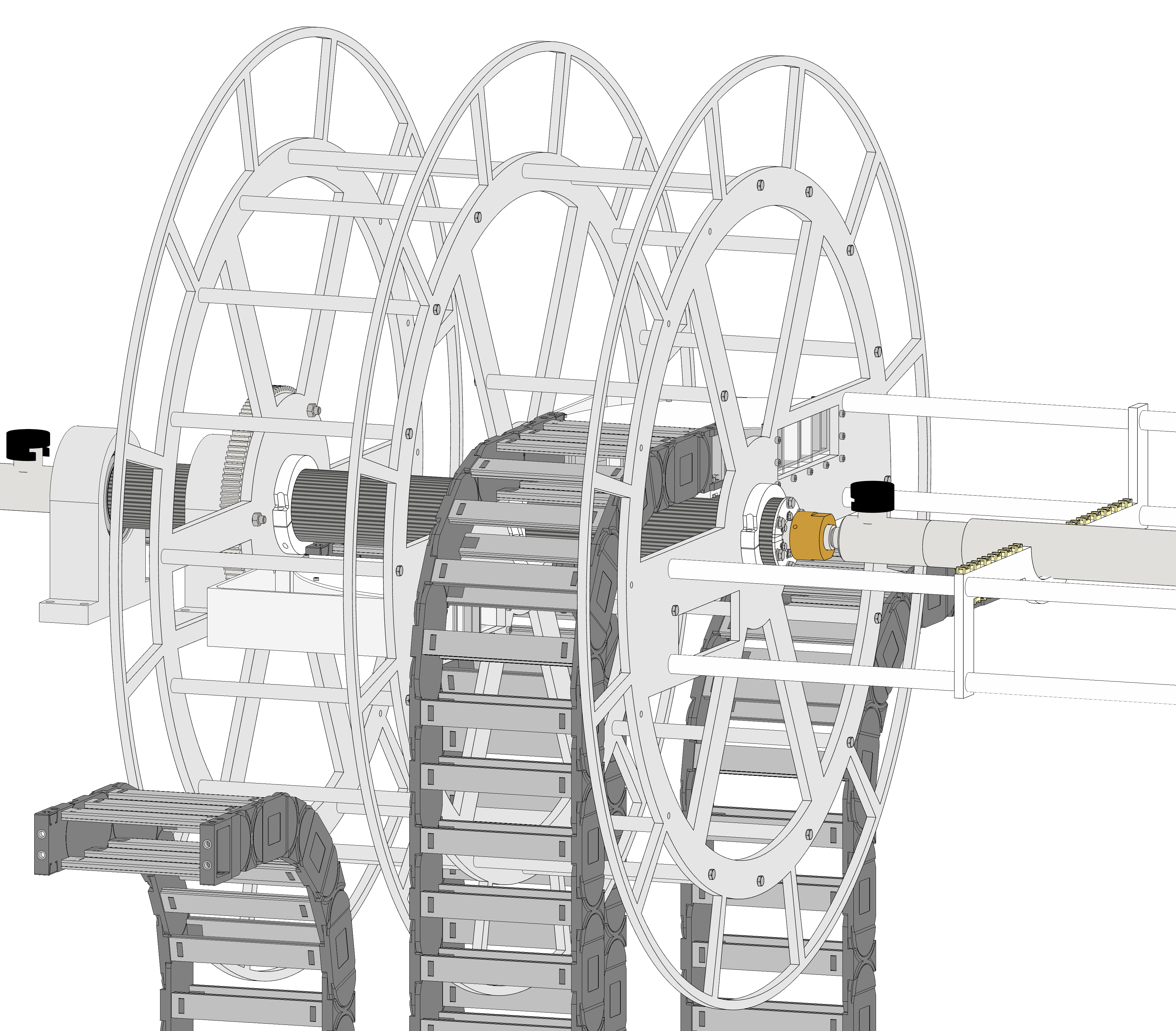}
\includegraphics[height=7.5cm]{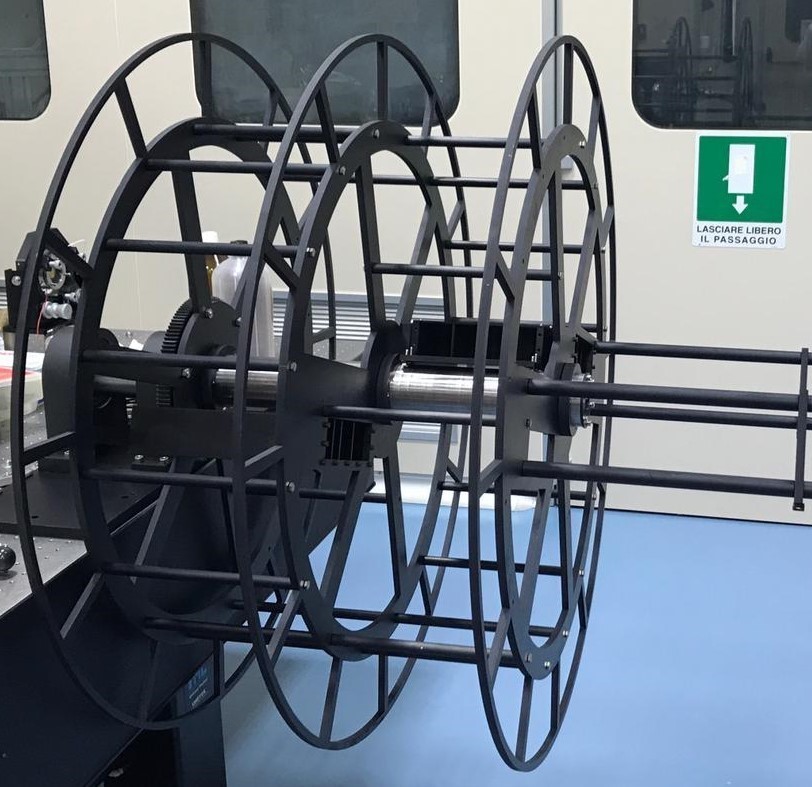}
\end{tabular}
\end{center}
\caption[example1] 
{ \label{fig:corot} 
The cable co-rotator support structure. Left: design; Right: as built (cable wrap missing in the picture).}
\end{figure}

\section{INTRODUCTION}
\label{sec:intro}  
The transients are astronomical events that last only for a limited time. In the last decades, the research on transients has generated many of the most impacting discoveries in astrophysics, like the gamma-ray bursts, the accelerating universe, the super-luminous supernovae and, more recently, the electromagnetic counterparts of gravitational wave events. Moreover, the future decade is expected to be a golden age for the study of transients, with new extremely powerful survey facilities coming into operation. As of today, the transients are being discovered at a rapid rate, but the number of discoveries will enormously increase with the future survey telescopes like the Vera Rubin telescope. The space for new discoveries is immense, provided that the new discovered objects are rapidly followed up by dedicated optical/near-infrared telescopes with spectroscopic capabilities. 

Nevertheless, so far most of the transient objects discovered by the current surveys remain unclassified, as the observing time usually granted for their classification and follow-up is largely insufficient. Thus, it is expected that when the future large scale survey machines will start operations, if no actions are taken, the bottleneck on the spectroscopic follow-up will be unsustainable. 

SOXS is designed exactly to follow up rapidly the new discovered objects, simultaneously at optical and near-infrared wavelengths. 
Thus, the science case of the instrument is combined with a good timing, which makes it coming into operation in parallel with many synergic facilities: just to name a few, LSST, Euclid, Gaia, PanSTARRS, Zwicky Transient Factory for optical searches, Swift, Fermi, SVOM, MAGIC and CTA for high-energy objects and, in the multi-messenger domain, aLIGO/VIRGO for gravitational waves, and KM3NET and ICECUBE for neutrinos. 

SOXS will have a significant amount of observing time (the consortium will be granted with 900 NTT nights over 5 years) to classify the astrophysical transients and characterize and follow-up the most interesting objects. Meanwhile, the consortium has established dedicated working groups to prepare the forthcoming science phase. Their interests span over the classification and study of all kind of transients, e.g. supernovae, electromagnetic counterparts of gravitational wave events, neutrino events, tidal disruptions of stars in the gravitational field of supermassive black holes, gamma-ray and fast radio bursts, X-ray binaries and novae, magnetars, but also asteroids and comets, activity in young stellar objects, blazars, and AGN. 

The SOXS focussed science case perfectly fits into the strategy of specializing the two medium-class telescopes operated by ESO in the La Silla Observatory to dedicated tasks: the NTT with SOXS is devoted to the transients, whereas the 3.6-m is dedicated to exoplanets\cite{deZeeuw16}.

Apart from the realization of the instrument, the consortium is committed to handle the operation phase, managing the schedule for the consortium and the ESO community. The schedule will be flexible and updated daily in order to respond to fast alerts, with scientists on duty, ready to react. The consortium will also provide to the community essential tools, like the Exposure Time Calculator and the reduction pipeline. The operations will be mainly managed  remotely, but still with a telescope operator on site for the night duties.

\section{Design overview}
\label{sec:Design}  

Hereinafter, we summarize the instrument design, complementing the descriptions presented at previous conferences\cite{Schipani16,Schipani18} with the most recent updates, and reporting on the status of the project.

SOXS is a medium resolution spectrograph with an average  R$\sim$4500 for a 1 arcsec slit, capable of simultaneously observing over the spectral range 0.35-2.0$\mu$m. It follows the fundational concepts of the X-shooter\cite{Vernet11} at the ESO-VLT, albeit with a different instrument design. Obviously, for most of the time SOXS will work in spectroscopic mode, but the instrument offers imaging capabilities as well through the acquisition camera, allowing for multi-band photometry of the faintest transients in the optical band (ugrizY + V Johnson). 

A 3D layout of SOXS at the Nasmyth focus of the NTT is shown in Fig.~\ref{fig:SOXS}. Most of the parts in the figure have been  manufactured and are currently available at the various SOXS laboratories.

The instrument is composed of two distinct spectrographs, for the UltraViolet-Visible (350-850 nm) and the Near InfraRed (800-2000 nm) bands. The wavelength overlap has been designed to allow for the cross-calibration of the two arms. The UV-VIS arm is based on a novel multi-grating concept\cite{Rubin18}, imaging different narrow band spectra on a single ``wonder" camera\cite{Oliva16}, whereas the NIR arm\cite{Sanchez18} implements a layout with collimator compensation of camera chromatism\cite{Delabre89}. 

The two arms are connected to a common opto-mechanical system, the ``Common Path", which splits the light from the telescope focus to the two spectrograph slits, through relay optics which reduce the F/number from F/11 to F/6.5 and compensate for the atmospheric dispersion in the UV-VIS arm. Additionally, the Common Path drives the light to/from the other instrument sub-systems, i.e. the $3.5'\times 3.5'$ acquisition and guiding camera and the unit for the wavelength and flux calibration. 

The functional diagram of the instrument is represented in Fig.~\ref{fig:control}.

\section{Status of instrument realization}
\label{sec:status} 

\subsection{The structure}
At the NTT, the Nasmyth platforms at the two foci are not an integral part of the telescope, they are rather built for the necessities of the instruments. Following a trade-off analysis between maintaining the existing Nasmyth platform or not, we decided to replace it with a brand new one, tailored for SOXS. It hosts the two electronic cabinets and a cable co-rotator to drive cables and pipes through two cable-wraps.

A telescope simulator loaned by ESO, replicating exactly the NTT mechanical interface and provided with a rotating bearing, has been moved to the INAF laboratories for testing the instrument in any gravity condition. During the pre-integration activity in Europe, the new platform will not be used because the telescope simulator cannot accomodate it. The structural parts that will be needed during the pre-integration in Europe, i.e. the interface flange and the co-rotator system, have been manufactured, pre-mounted and checked. A preliminary alignment of the sub-system interfaces to the flange has also been done. 

The co-rotator is driven by a servo motor that follows the analogue signal coming from linear sensors pushed during the instrument rotation. The electronics has been validated and is going to be integrated with the mechanics.

Figure~\ref{fig:flange} shows the interface flange with the co-rotator sensing system, whereas Fig.~\ref{fig:corot} shows the co-rotator structure that will host the two cable-wraps.

\begin{figure} [ht]
\begin{center}
\begin{tabular}{c} 
\includegraphics[height=8cm]{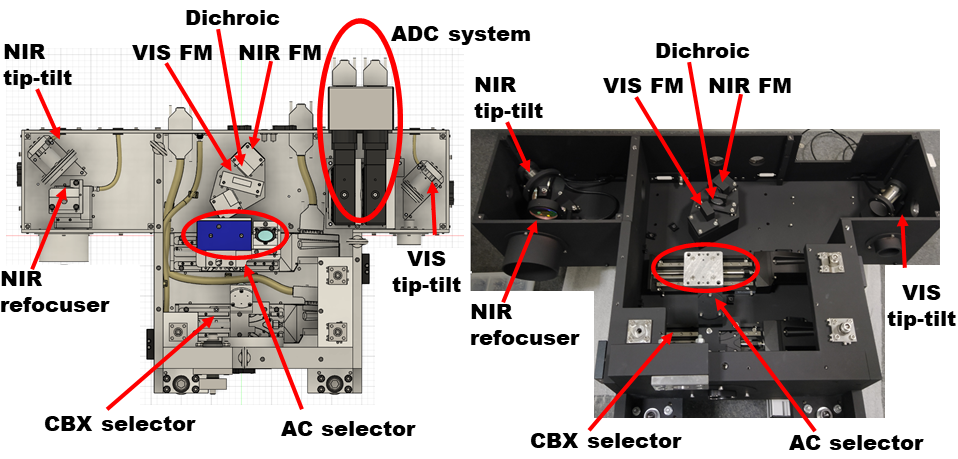}
\end{tabular}
\end{center}
\caption[example1] 
{ \label{fig:CP} 
The Common Path unit. Left: design; Right: real system during integration activities. The ADC system and the Acquisition Camera mirror are missing in the picture.}
\end{figure}

\subsection{Common Path}
Figure~\ref{fig:CP} shows the Common Path\cite{Claudi18} in its advanced status of realization. In normal operations, the light coming from the telescope through an instrument shutter is split by a dichroic, which reflects the visible and transmits the infrared wavelengths to two folding mirrors, deviating the beams to tip-tilt devices for the active compensation of the instrument flexures. The atmospheric dispersion is corrected by an ADC in the visible arm, where the atmospheric effect is more severe. The infrared arm is provided with a focusing mechanism which, together with the telescope secondary mirror axial displacement, gives the two degrees of freedom to get always the best focus on both arms.

In calibration mode, a selector mirror on a movable slit allows us to pick the light from the calibration lamps. A further selector device allows us to redirect the light to the acquisition camera, for acquisition of the target or light imaging. 

The Common Path structure was manufactured and underwent the interface test with the flange successfully. The procurement of the optical components is almost completed, with the only exception of the Atmospheric Dispersion Corrector assembly, composed of two quadruplets installed in counter-rotating devices. All the other optical elements are available, glued on their mounts and installed on the optical bench, where the preliminary optical alignment has been completed\cite{Biondi20}. The actuator control loops have been preliminary tuned and the instrument software is ready for integrated tests.

\begin{figure} [ht]
\begin{center}
\begin{tabular}{c}
\includegraphics[height=6.6cm]{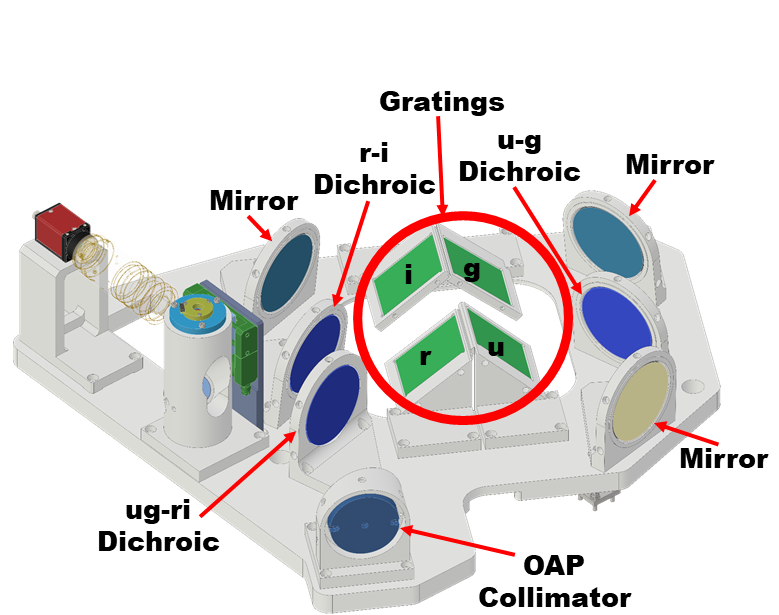}
\includegraphics[height=7.7cm]{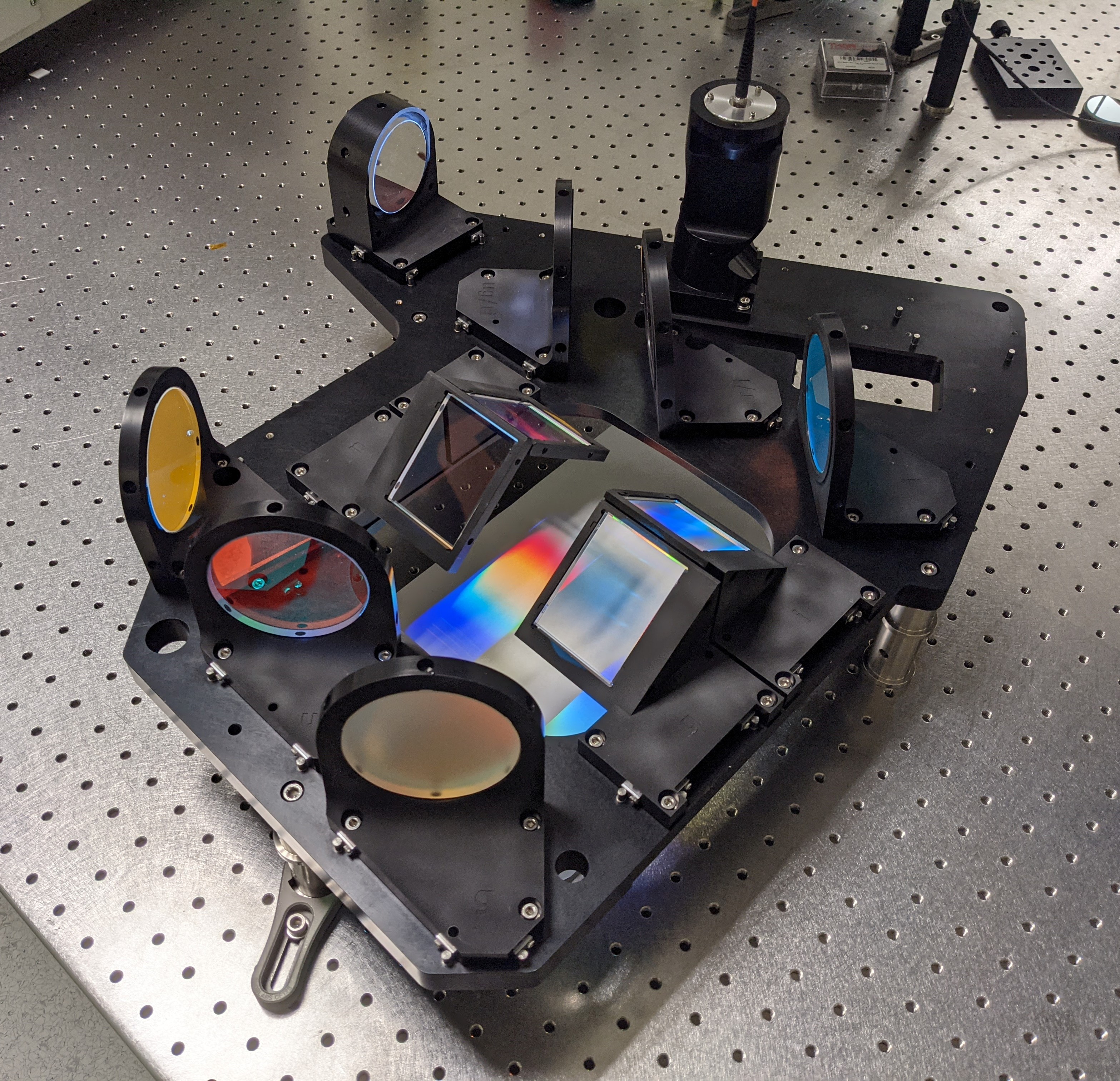}
\end{tabular}
\end{center}
\caption[example1] 
{ \label{fig:FeedMount} 
The UV-VIS spectrograph feed system. Left: design; Right: as built.}
\end{figure} 

\begin{figure} [ht]
\begin{center}
\begin{tabular}{c} 
\includegraphics[height=7.5cm]{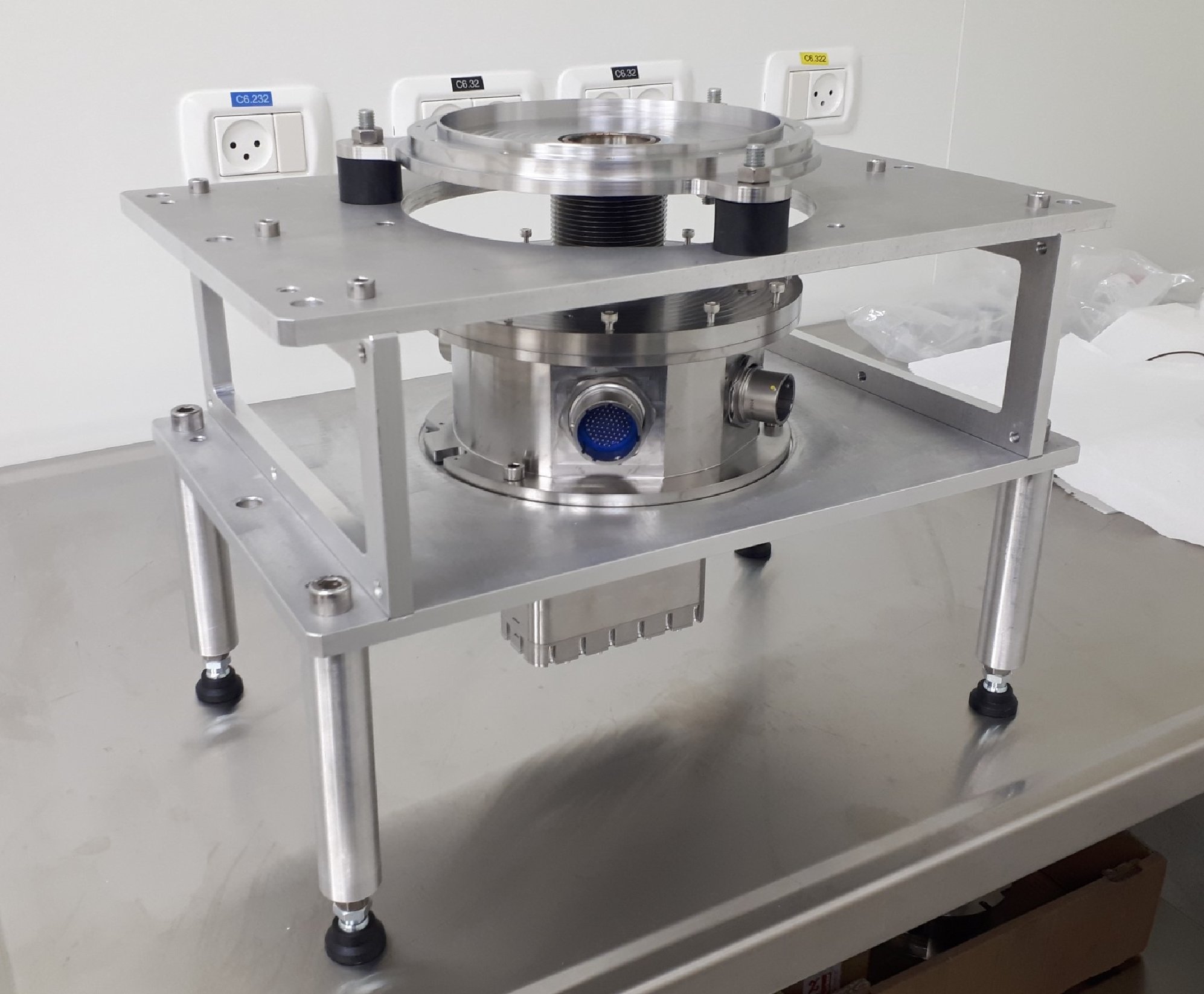}
\includegraphics[height=7.5cm]{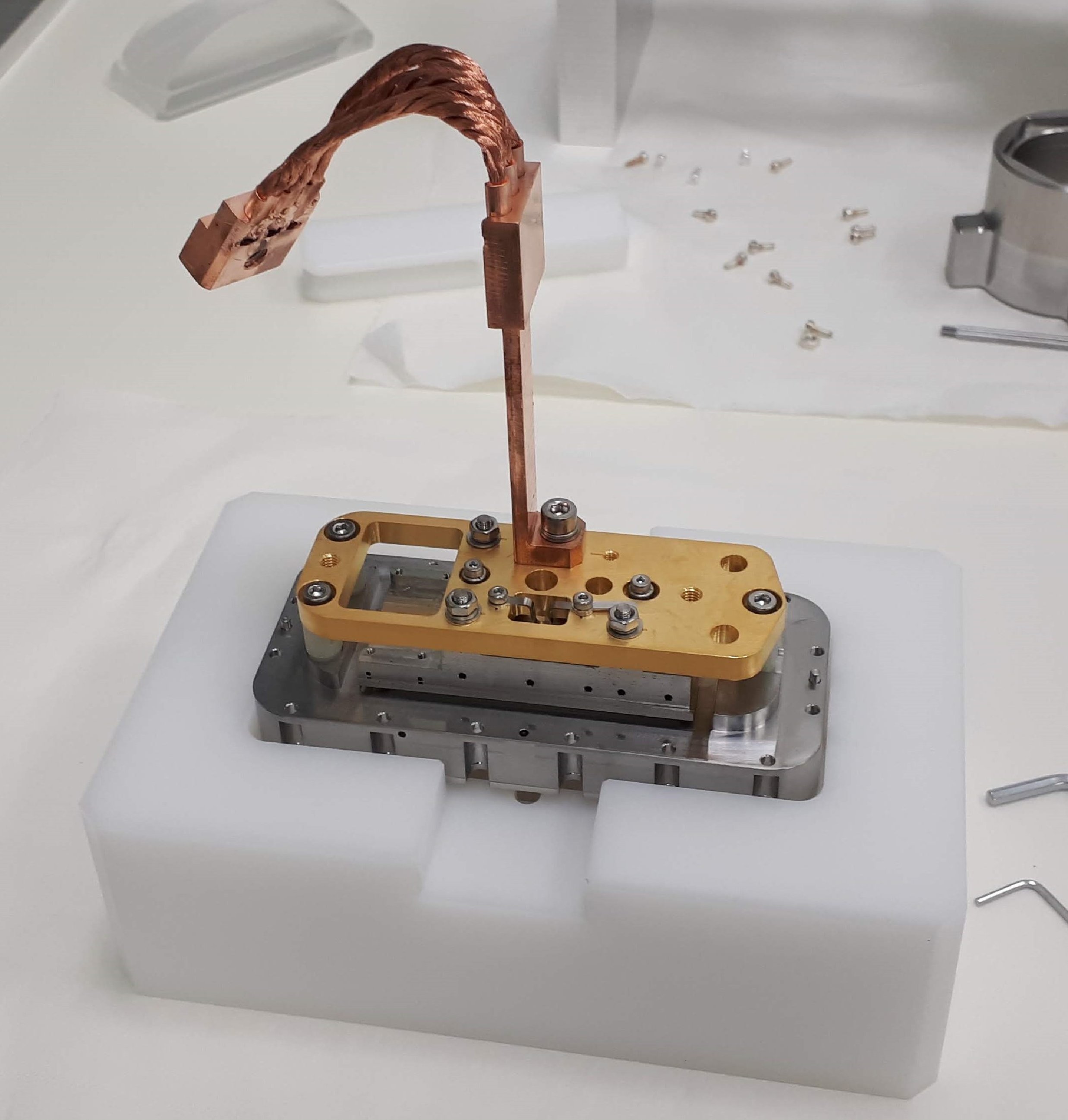}
\end{tabular}
\end{center}
\caption[example1] 
{ \label{fig:CCDChamber} 
The UV-VIS spectrograph CCD dewar and mount.}
\end{figure}

\subsection{UV-VIS Spectrograph}

The UV-VIS spectrograph\cite{Rubin18} can be divided in two optical sub-systems. 

The first part is a feed which collimates the beam coming from the Common Path, splits the beam to four, each beam with $\Delta\lambda\sim$100 nm, and disperses each beam using a novel design based on multiple ion‐etched gratings optimized for the relevant beam waveband. The feed optics are an off-axis parabola, acting as collimator, and a set of dichroics, folding mirrors and gratings. Their procurement has been completed and all the parts are in specs. The feed system optomechanics is ready (Fig.~\ref{fig:FeedMount}) and the alignment of the optics is well on the way with good results. 

The second part is a Schmidt camera fed by the four beams simultaneously, designed to be very compact  with one mirror and two lenses, where the field flattener is the CCD window. The camera optics are going to be delivered and their installation is envisaged immediately after. 

Meanwhile, the detector system\cite{Cosentino18} is being assembled. It is based on a e2v CCD 44-82, a custom detector head coupled with a Continuous Flow Cryostat (CFC) cooling system and the NGC CCD controller developed by ESO. The CCD chamber (Fig.~\ref{fig:CCDChamber}) is ready and tests of the detector installation procedure have been performed. The detector system has been procured and is currently under test. 

The integration of the opto-mechanics with the detector system and the validation of the whole unit are forthcoming.

\begin{figure} [ht]
\begin{center}
\begin{tabular}{c} 
\includegraphics[height=9cm]{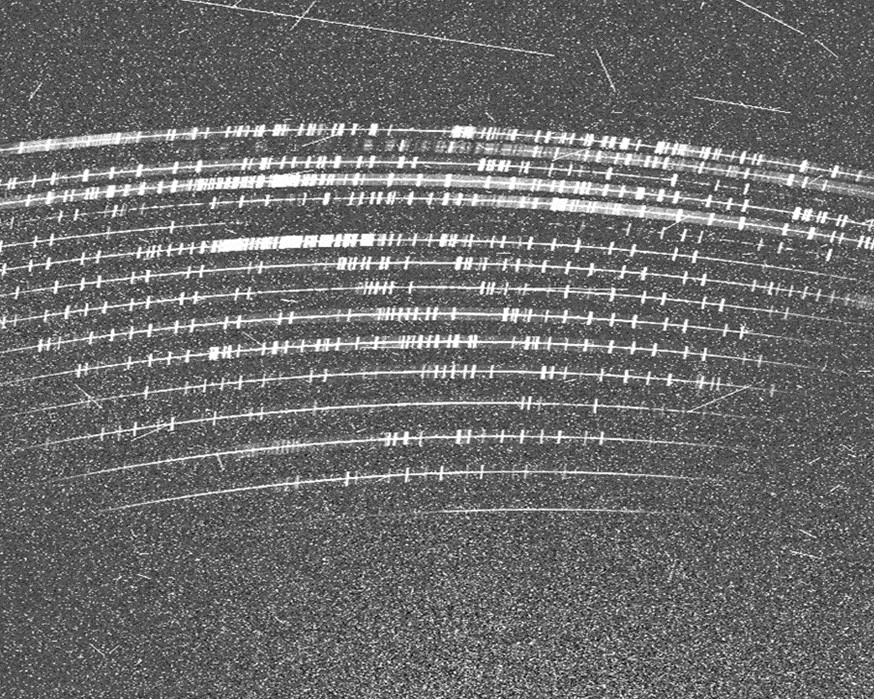}
\end{tabular}
\end{center}
\caption[example1] 
{ \label{fig:synt} 
NIR arm synthetic frame.}
\end{figure} 

\begin{figure} [ht]
\begin{center}
\begin{tabular}{c} 
\includegraphics[height=4.4cm]{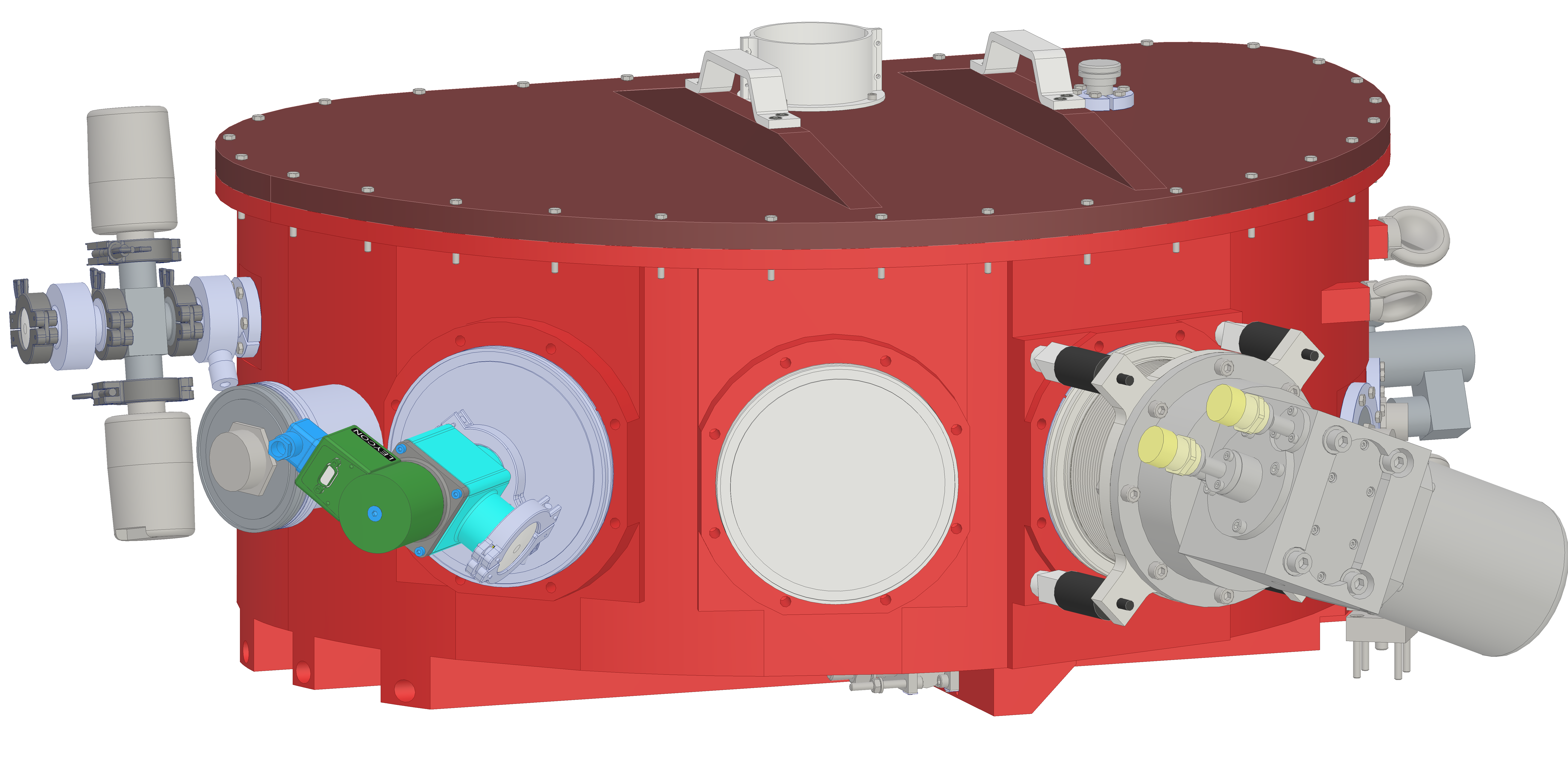}
\includegraphics[height=4.4cm]{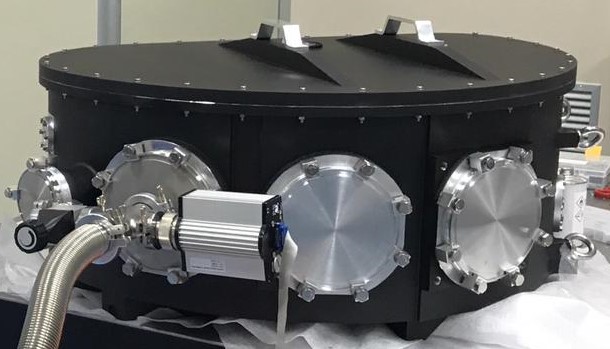}
\end{tabular}
\end{center}
\caption[example1] 
{ \label{fig:vv} 
The NIR spectrograph vacuum vessel. Left: design; Right: as built.}
\end{figure}

\subsection{NIR Spectrograph}
The NIR spectrograph\cite{Vitali18} is a fully cryogenic echelle-dispersed spectrograph, working in the range 0.80-2.00 $\mu$m, equipped with an 2k x 2k Hawaii H2RG IR array from Teledyne, working at 40K. 

The optical design is composed of a double pass collimator and a refractive camera; the dispersion is obtained via a main disperser grating and three prisms acting as cross-dispersers. The collimator receives a F/6.5 input from the Common Path and creates a collimated beam. The main disperser is a standard grating with 72 l/mm and a blaze angle of $44 {\deg}$, whereas three Cleartran prisms, used in double pass, provide the cross-dispersion. The camera, a completely transmissive system composed of three single lenses, re-images on the detector the focal plane produced by the second pass through the collimator. The spectrum is cross-dispersed on 15 orders: Fig.~\ref{fig:synt} shows a synthetic spectrum obtained through the SOXS instrument simulator\cite{Genoni20}.

The spectrograph is cooled down to $\sim$150 K to lower the thermal background and is equipped with a thermal filter to block any thermal radiation above 2.0$\mu$m. The cryogenics is operated via a Closed-Cycle Cryocooler.

Figure~\ref{fig:vv} represents the D-shaped vacuum vessel, recently delivered, that will be interfaced to the flange through a set of kinematic mounts. Most of the NIR spectrograph optical elements are available in SOXS laboratories; the procurement is soon expected to be over. Afterward, the sub-system will be fully integrated and tested.  

\subsection{Acquisition Camera}
The Acquisition Camera\cite{Brucalassi18} is used for target acquisition, (optional) secondary guiding and for providing scientific photometric observations.

The system consists of a collimator lens, a folding mirror, a filter wheel equipped with a broad-band filter set (ugrizY and V-Johnson), focal reducer optics, and a CCD camera, included in an aluminum structure. The detector is a 1k x 1k Andor iKon-M 934 camera. The driver to run the camera under the SOXS control software has been developed and validated with the real system.

The sub-system receives a F/11 beam from the telescope through a selector folding mirror placed in the Common Path, having different positions for different functions. 

In acquisition and imaging mode, the selector mirror redirects the full field towards the camera. 

In spectroscopy mode, the selector mirror passes through a hole an unvignetted field of 15 arcsec to the spectrograph slits, whereas the peripheral field is simultaneously imaged on the camera. Thus, a secondary guiding\cite{Ricci18} is possible using peripheral sources.

At the time of writing, all orders related to the camera have been placed, many system components are available and the remaining parts are going to be delivered soon, prior to the final integration phase.

\begin{figure} [ht]
\begin{center}
\begin{tabular}{c} 
\includegraphics[height=5.3cm]{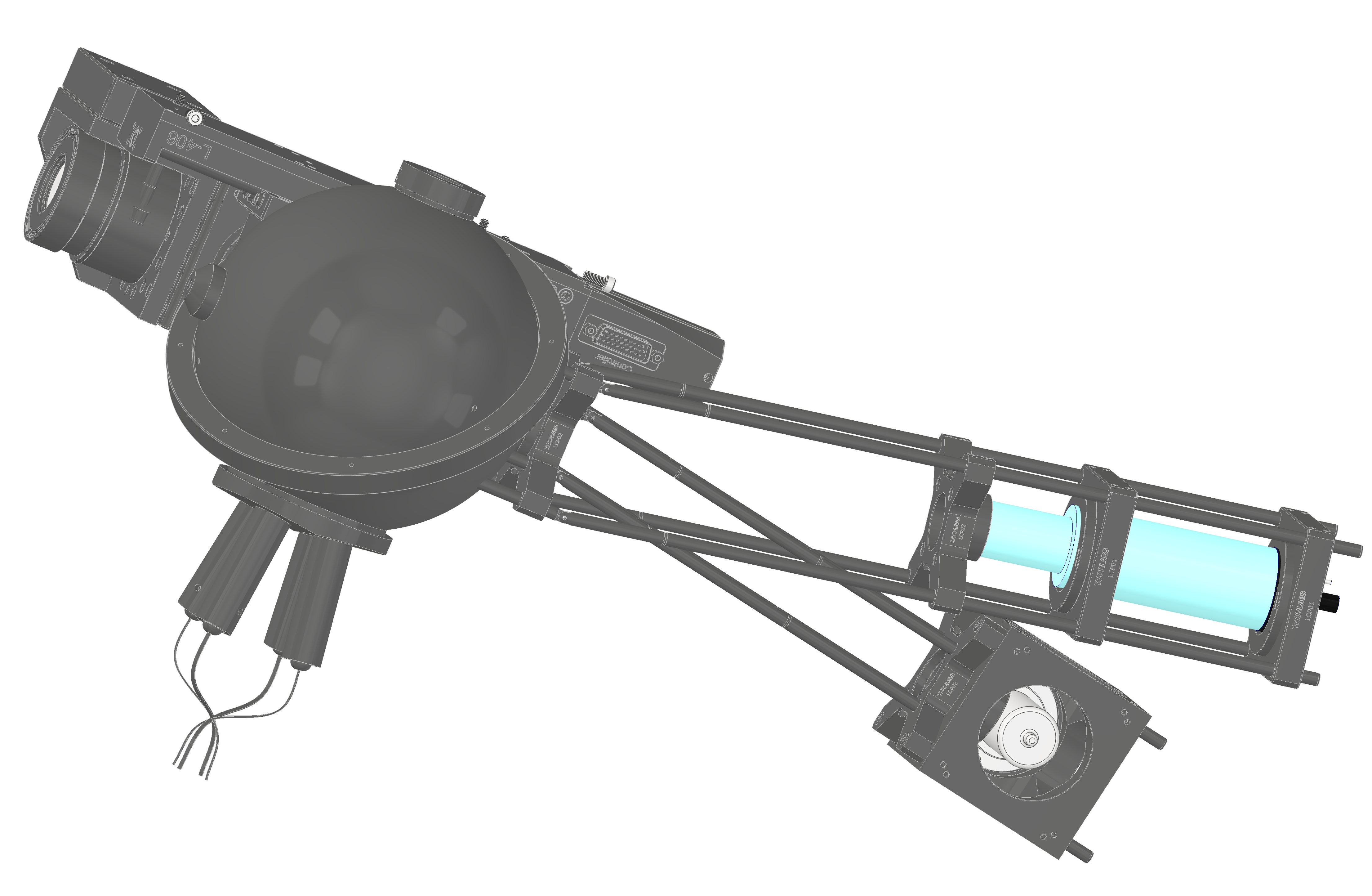}
\includegraphics[height=5.3cm]{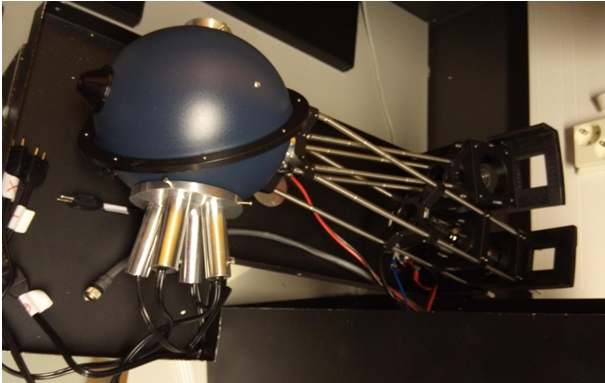}
\end{tabular}
\end{center}
\caption[example1] 
{ \label{fig:cbx} 
Calibration Unit. Left: design; Right: as built.}
\end{figure}

\subsection{Calibration Unit}
Figure~\ref{fig:cbx} represents the Calibration Unit\cite{Kuncarayakti20} design (left) and a picture of the system integration (right). The light is emitted out of the exit port of an integration sphere. The manufacturing of the unit is basically completed, thus system tests are upcoming.

On the left side of the unit, the integration sphere with the 4 Ne-Ar-Hg-Xe pen-ray lamps bundled together for NIR wavelength calibration is visible. The individual lamps are controlled to operate together as one lamp. The mount of the Quartz-Tungsten-Halogen (QTH) lamp, for flux calibration in the 500-2000 nm range, is on the opposite side of the sphere. On the right side of the unit, the Th-Ar hollow cathode lamp (cyan in the drawing) for UV-VIS wavelength calibration and the Deuterium (D2) lamp, for flux calibration in the 350-500 nm range (used simultaneously with QTH lamp for UV-VIS arm flux calibration), are visible as well.

The light coming from the unit goes through relay optics to get a uniform illumination of the spectrograph slits, replacing the light coming from the telescope. The selection between the telescope and the calibration unit beams is implemented by a moving mirror in the Common Path.

The unit includes a synthetic star mode based on a pin-hole mask, which can be placed next to the integrating sphere, and lenses to re-image it at the telescope focal plane. It is used to simulate an artificial star for engineering purposes.

\begin{figure} [ht]
\begin{center}
\begin{tabular}{c} 
\includegraphics[height=7.5cm]{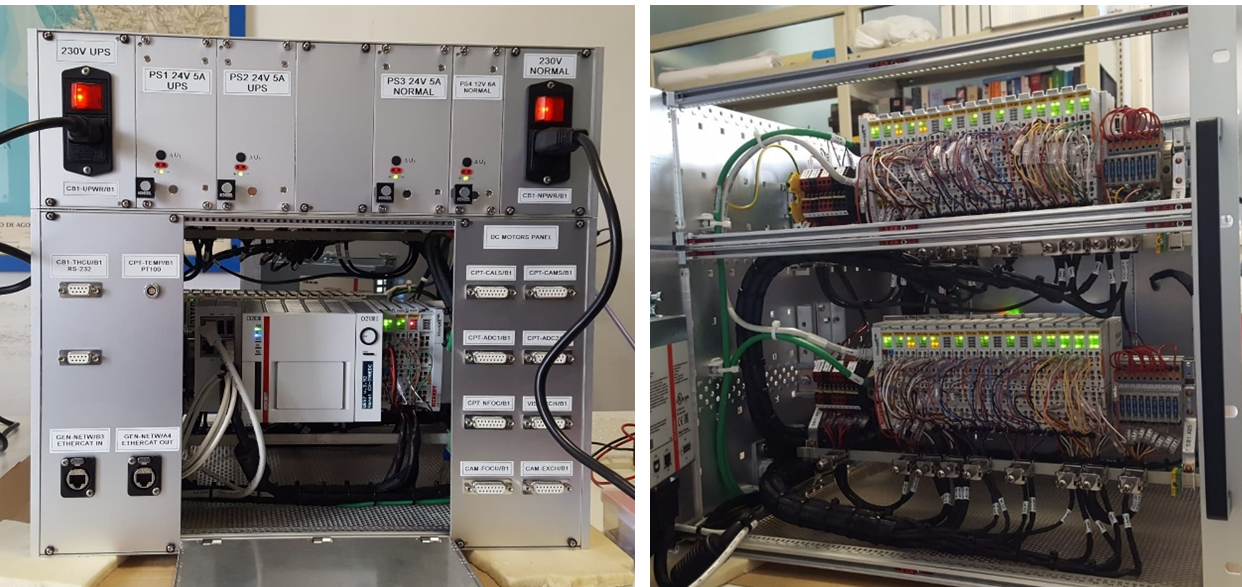}
\end{tabular}
\end{center}
\caption[example1] 
{ \label{fig:ele} 
Control electronics under test.}
\end{figure} 

\begin{figure} [ht]
\begin{center}
\begin{tabular}{c} 
\includegraphics[height=7.5cm]{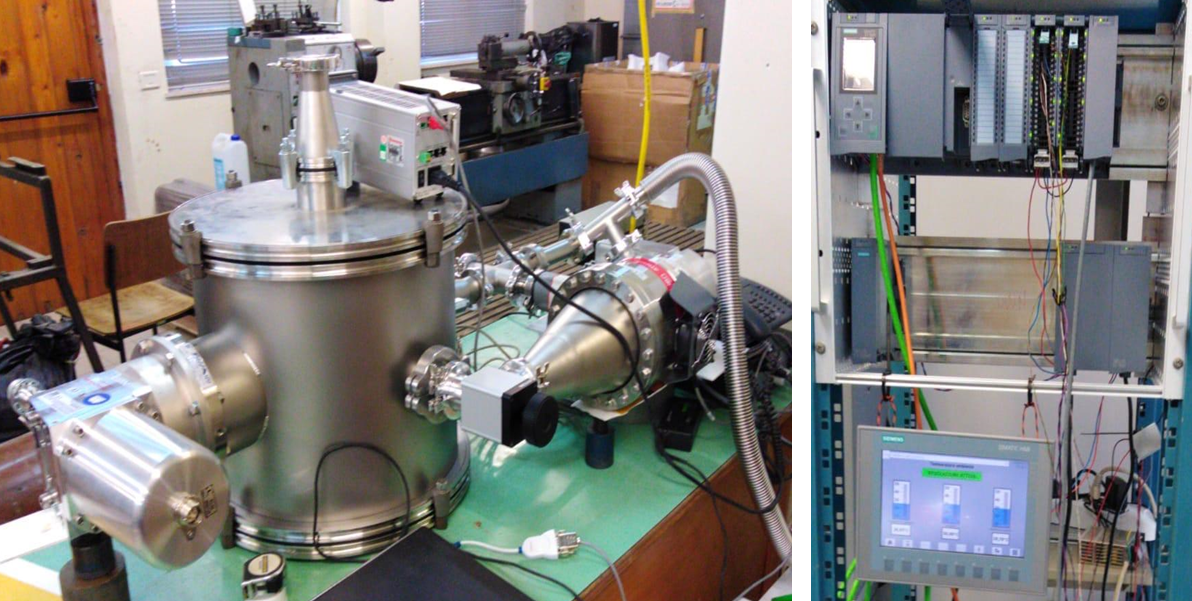}
\end{tabular}
\end{center}
\caption[example1] 
{ \label{fig:cryo} 
Cryogenics system under test with a dummy vessel in the lab.}
\end{figure} 

\begin{figure} [ht]
\begin{center}
\begin{tabular}{c} 
\includegraphics[height=10cm]{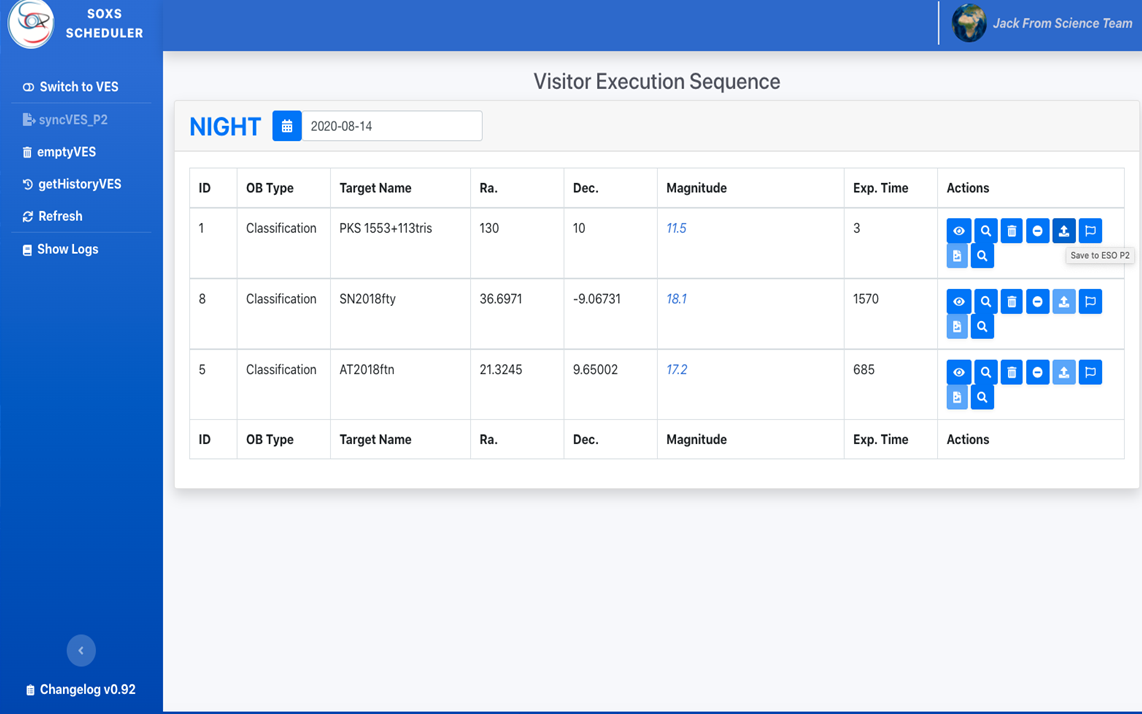}
\end{tabular}
\end{center}
\caption[example1] 
{ \label{fig:sched} 
The SOXS Scheduler App.}
\end{figure} 

\subsection{Instrument Control}
The Instrument Control Electronics\cite{Capasso18} is based on a  Beckhoff PLC to control the instrument functions. The PLC is equipped with modules to interface with motor drives, encoders, and all the other hardware devices. The architecture is based on an EtherCAT fieldbus network, which allows us to distribute the control modules, wherever they are needed. 

In the case of SOXS, this allowed us to use only one CPU, communicating through the fieldbus network with local control modules physically located in different sub-racks and cabinets. 

An Open Platform Communication - Unified Architecture (OPC-UA) server is installed on the PLC: it stores all the process variables accessible to the Instrument Software\cite{Ricci18}, based on the VLT Common Software layer and running on an Instrument Workstation. The Linux Instrument Workstation adopts the OPC-UA protocol to communicate with the PLC.

This architecture follows the ESO standards, favouring the integration and maintenance of the instrument within an ESO observatory. The effectiveness of these solutions has been proved by recent instruments like ESPRESSO\cite{Baldini20}.

Figure~\ref{fig:ele} shows one of the 19-inch sub-racks hosted in the two electronic cabinets. Most of the software and control electronics is ready\cite{Colapietro20}, and tests with the instrument are forthcoming.

Figure~\ref{fig:cryo} shows the laboratory activities related to the Cryo-Vacuum system. This is a fully independent sub-system, by the control point of view. It is based on a Siemens PLC that must run continuously, reliably and independently of any other instrument necessity. Thus, the main instrument software does not control the Cryo-Vacuum system status, but just reads its status.

\section{Preparing the operation phase}
\label{sec:opera} 

\subsection{The Scheduler}
With SOXS coming into operation, the way the NTT is operated will change a lot. So far, the operations are fully handled by ESO. The astronomers submit proposals twice a year, the ESO Observing Programmes Committee (OPC) panels review, accept and give them a grade. Then, the Principal Investigators of the programmes prepare and submit the corresponding Observation Blocks through the P2 web application, ESO prepares the schedule and finally the observations are performed in Visitor Mode, i.e. with astronomers traveling to Chile. The astronomers are assisted in the whole process by an ESO department. 

This ``standard" paradigm will change significantly with SOXS. The proposals will still be evaluated by the OPC, but the responsibility of the operations will go to the consortium, that consequently will provide services to the entire users' community. Also, the transient science requires a heavy usage of Target of Opportunity mode, i.e. the targets are not known in advance and the schedule cannot be static. Thus, the consortium will work continuously on the selection of the targets, the generation of the Observation Blocks, the scheduling of the nights, mixing the targets coming from the consortium and the community. 

In the organization of this transition, as most of the consortium people do not reside in Chile, one guideline has been the ability to govern remotely most of the new process.

A new framework is under development, where the consortium representatives will be assisted by a software application to select the targets, generate the Observation Blocks, schedule the night. This application is going to be a fundamental tool, given the enormous discovery rate of transients expected in the next decade from the Rubin telescope and the other facilities. 

The target selection will benefit from the heritage of the PESSTO (recently renamed to ePESSTO+) spectroscopic survey, in execution at the NTT with the existing spectrographs since several years. The PESSTO Marshall application, fed with transients discovered by the existing surveys (e.g. ATLAS, ZTF, etc.), is a powerful aid for the time-domain astronomers to select the transients for classification and follow-up. It talks directly to the survey databases, assimilating all information about the targets, and recording classifications and all available data in its own database\cite{Smartt13}.

This concept will be evolved to a SOXS Marshall, becoming part of a more complex system that will interface, on the other side, to the ESO existing infrastructure\cite{Bierwirth18}. The SOXS software will generate automatically the Observation Blocks through the P2 API, pushing them daily into the night execution sequence at the telescope. The application will take care of the observability of the targets and weather conditions, retrieving data from the observatory Astronomical Site Monitor system. In case any weather or technical issue affects the envisaged execution sequence, a new schedule will be regenerated on the fly, optimized for the new conditions and the remaining part of the night. 

This scheduler software\cite{Landoni20} (under development, see Fig.~\ref{fig:sched}) is characterized by a high available and scalable architecture, implementing state-of-the-art technologies for API application like Docker Container, API Gateway and Python-based Flask framework.

\subsection{The Pipeline}
SOXS data science and data quality control products will be similar to those produced for X-shooter, but with the important difference that science ready reduced spectra shall be produced immediately after an exposure finishes, due to the focus on rapid response to transient objects.

The reduction tools of many instruments produce quick-look data within seconds of an exposure finishing, but science ready data are ready within days to weeks: to achieve SOXS science goals, delays in accessing fully reduced, science ready products, are not acceptable. Thus, science ready, fully reduced spectra will be produced immediately after an exposure has finished. 

The stability of calibration frames (bias, darks, flat-fields, arc calibrations and flux calibration) is the key for producing rapid data products which are scientific grade. 

The SOXS pipeline\cite{Young20} development is well underway, currently focused on the capability of reducing X‐shooter data. The pipeline is being written in Python 3, using many mature open‐source packages such as Numpy, CCDProc, Matplotlib, and other Astropy affiliated projects. It will be available as a service for the entire SOXS users' community.

\section{Conclusions}
\label{sec:concl}
SOXS is going to be the work-horse of ESO observatories for the spectroscopic follow-up of transient sources. All the sub-systems are in an advanced phase of realization, very close to be completed and shipped to the preliminary integration site in Italy. The instrument integration and test phase is imminent and planned in 2021. 

Needless to say, the impact of the pandemic on a project in manufacturing and integration phase is severe, thus SOXS is experiencing delays. Currently making plans is harder than usual; however, at the time of writing, SOXS is planned to be on sky within 2022.

\bibliography{SOXS} 
\bibliographystyle{spiebib} 

\end{document}